\documentclass[12pt]{article} 
\usepackage{amssymb}
\usepackage{graphicx}
\begin{document} 
\begin{center}
{\large \bf Testing scaling laws for the elastic scattering of protons}

\vspace{0.5cm}                   

{\bf I.M. Dremin, V.A. Nechitailo}

\vspace{0.5cm}                       

         Lebedev Physical Institute, Moscow 119991, Russia\\

\end{center}

\begin{abstract}
Theoretical proposals of scaling laws for the differential elastic scattering 
cross sections of protons are confronted with experimental data over a
wide energy range. Different combinations of the transferred momentum and 
energy resulting from the solution of the definite partial differential equation
are attempted as scaling variables. Reasonable scaling of the differential 
cross sections in the diffraction cone has been shown for one of these 
variables. The violation of the geometrical scaling is ascribed to the increase
of the proton blackness with energy.
The origin of high-t region violations of scaling laws is discussed.
\end{abstract}

The differential cross section for elastic scattering of particles
$d\sigma (s,t)/dt$ is the only measurable characteristics of this process.
At any fixed energy $s$, one presents a one-dimensional plot of its dependence
on the transferred momentum $t$. However, the possibility that the differential 
cross sections might be described as functions of a single scaling variable 
representing a definite combination of energy and transferred momentum has 
been discussed \cite{akm71, ddd}. No rigorous proof of this
assumption has been proposed. Recently this property was obtained \cite{drad}
from the solution of the partial differential equation for the imaginary part
${\rm Im}A(s,t)$ of the elastic scattering amplitude. The equation has been 
derived by equating the two expressions for the ratio of the real to imaginary
parts of the amplitude $\rho (s,t)$. They were known from the local dispersion 
relations \cite{gmig, sukha, fkol} with the $s$-derivative and from the
linear $t$-approximation \cite{akm71, mar1} with the $t$-derivative. These
expressions are, correspondingly,
\begin{equation}
\rho (s,t)= \frac {\pi }{2}
\left [\frac {\partial \ln {\rm Im}A(s,t) }{\partial \ln s }-1\right ]
\label{rhodit}
\end{equation}
and
\begin{equation}
\rho (s,t)=\rho (s,0) \left [1+\frac {\partial \ln {\rm Im}A(s,t) }
{\partial \ln \vert t\vert }\right ].
\label{rhotau}
\end{equation}
Therefrom the following partial differential equation is valid
\begin{equation}
p-f(x)q=1+f(x),
\label{partial}
\end{equation}
where $p=\partial u/\partial x ; \; q=\partial u/\partial y; \;
u=\ln {\rm Im}A(s,t); \; f(x)=2\rho (s,0)/\pi \approx d\ln \sigma _t/dx ; 
\; x=\ln s; \; y=\ln \vert t\vert ; \; \sigma _t$ is the total cross section. 
The variables $s$ and $\vert t\vert $ should be considered as scaled by the 
corresponding constant factors $s_0^{-1}$ and $\vert t_0\vert ^{-1}$.

Eq. (\ref{partial}) can be rewritten in another way as
\begin{equation}
\frac {\partial u}{\partial \ln \sigma _t}-\frac {\partial u}{\partial \ln t}=
1+\frac {d\ln s}{d\ln \sigma _t}.
\label{part1}
\end{equation}

The general solution of Eq. (\ref{part1}) reveals the scaling law
\begin{equation}
\frac {t}{s}{\rm Im}A(s,t)=\phi (t\sigma _t). 
\label{scal}
\end{equation}

For the differential cross section it looks like
\begin{equation}
t^2 d\sigma /dt=\phi ^2(t\sigma _t),
\label{sl1}
\end{equation}
if the real part of the amplitude is neglected compared to the imaginary part.
Thus the scaling law is predicted not for the differential cross section itself
but for its product to $t^2$.
Let us note that the often used ratio (see, e.g., \cite{bddd}) of $d\sigma /dt$
to $d\sigma /dt\vert _{t=0}\propto \sigma _t^2$ is also a scaling function.
However, expression (\ref{sl1}) is more suitable for comparison with 
experiment.

The scaling law with the $t\sigma _t$-scale is known as "geometrical scaling".
Different aspects of its violation are often discussed. Here we contribute
another view of this problem.

\begin{figure}[h]
%plot0_err.gnu
 \includegraphics[width=\textwidth]{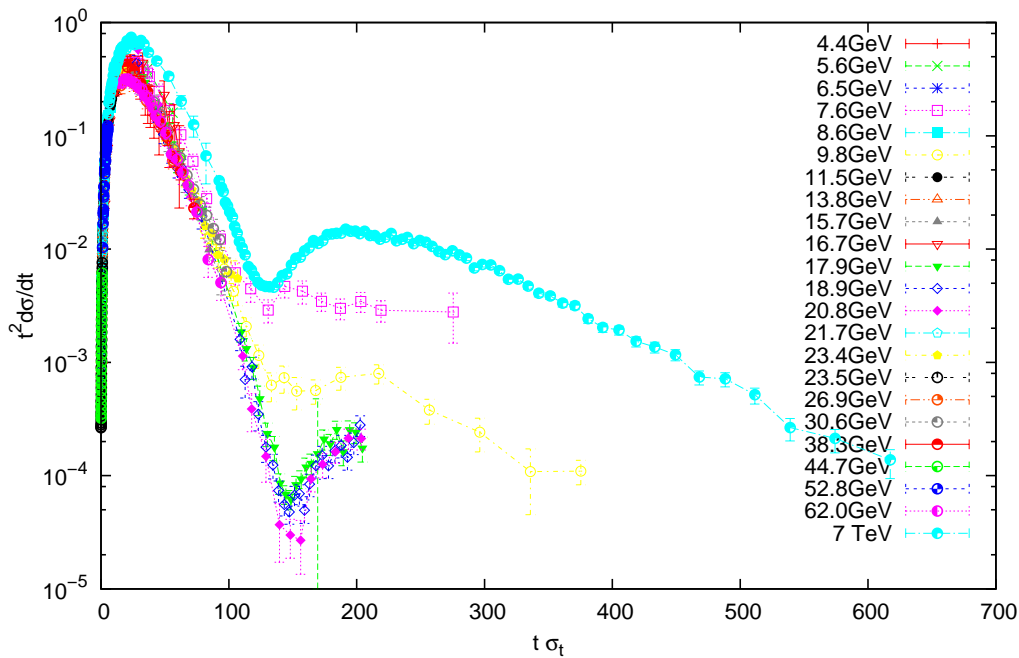}
 \caption{(Color online) The values of $t^2d\sigma /dt$ for pp-scattering at energies $\sqrt s$
from 4.4 GeV to 7 TeV as functions of $t\sigma _t$ with $\sigma _t$ 
provided by the corresponding experiment. The scale on the abscissa axis is 
defined by $t_0=-1$ GeV$^2$ and $\sigma _t$ in GeV$^{-2}$. The data are  
from \cite{cude, pdg, toteml, totemh}.}
 \label{fit_sigma_tot}
\end{figure}

The geometrical scaling violation is seen in Fig.~\ref{fit_sigma_tot} in the 
diffraction 
cone especially for the TOTEM data at 7 TeV. With the common approximation
$d\sigma /dt \propto \exp (Bt)$ in the diffraction cone one gets that the
maximum of the function $t^2d\sigma /dt$ displayed in Fig. 1 should be
positioned at $t_m\sigma _t=2\sigma _t/B=16\pi (1-\exp (-\Omega (s))$. 
It is important that it depends only on the opacity of protons $\Omega (s)$ 
(see the Table 1 in the review paper \cite{dr13}) but not on their radii. 
The shift of the maximum is completely determined by the energy increase of 
the opacity. Scaling is violated due to the stronger energy dependence of 
$\sigma _t$ compared to the diffraction cone slope $B$ observed in experiment.
That shows the difference between the volume effect important for the total
cross section and the surface effect responsible for the elastic processes
and, consequently, for the diffraction cone slope. Their energy dependences
coincide only if the opacity saturates.

Thus this simple geometrical scaling is not fulfilled at very high energies, 
even at low transferred momenta. We show below how this problem can be cured. 
Moreover, the scaling is much more strongly violated outside the diffraction region. 
It is also discussed in what follows.

In attempts to cure these problems we turn to the assumptions used in the 
derivation of the scaling law (\ref{sl1}). The neglect by the real part of 
the amplitude in (\ref{sl1}) is the most evident one. Its contribution is 
easily estimated using Eqs. (\ref{rhodit}), (\ref{rhotau}). One gets
\begin{equation}
t^2 d\sigma /dt=\phi ^2(t\sigma _t)[1+(d\ln \phi /d\ln (t\sigma _t))^2
\rho ^2(s,t=0)].
\label{sl2}
\end{equation}
The second term violates scaling - albeit it is very small because of
smallness of $\rho (s,t=0)$ and does not pose any problem. 

Another approximation is involved in the relation (\ref{rhodit}). It was 
guessed as the extension to non-zero transferred momenta of the first term
in the series expansion of the exact expression for $\rho (s,0)$ which looks
like
\begin{equation}
\rho (s,0)\approx \frac {1}{\sigma _t}\left [\tan \left (\frac {\pi }{2}
\frac {d}{d\ln s }\right )\right ]\sigma _t=
\frac {1}{\sigma _t}\left [ \frac {\pi }{2}\frac {d}{d\ln s }+
\frac {1}{3}\left (\frac {\pi }{2}\right )^3\frac {d^3}{d\ln s^3 }+...\right ]
\sigma _t.
\label{rhodi}
\end{equation}
The terms with higher derivatives in $s$ were neglected. This assumption is 
quite reasonable because their contribution seems negligible for experimentally 
measured energy dependence of $\sigma _t$ and to any analytical fits.

More serious questions arise concerning Eq. (\ref{rhotau}). It looks as if
only the first term in the $t$-expansion of $\rho (s,t)$ is taken into account 
in this relation. It could be satisfactory in the diffraction cone where
${\rm Im}A(s,t)\propto \exp (Bt/2)$. Let us note here that according to 
(\ref{rhotau}) $\rho (s,t)$ should become ever smaller in the diffraction cone 
crossing zero at $t=t_m$ and be negative at larger $\vert t\vert $. Moreover, 
even dealing within a linear approximation, one gets negative values of 
$\rho $ in the region directly attached to the diffraction cone (known as the
Orear region by the name of its discoverer) if $\rho (s,t)$ is approximated by 
its constant average value there \cite{drem}. 

The behavior of $\rho (s,t)$ may become there strongly non-linear 
in $t$ \cite{drem}. The solution of the unitarity 
equation for the imaginary part of the amplitude in the Orear region 
\cite{anddre, anddre1} (see also the review paper \cite{dr13}) 
$u\propto -[2B\ln (4\pi B/\sigma _tf_{\rho })\vert t\vert]^{0.5}$
is quite complicated and does not seem to satisfy the scaling law. Here,
$f_{\rho }=1+\rho (s,0)\rho _l$ with
$\rho _l$ denoting the average value of $\rho $ in this region. If $\rho _l$
is replaced by the non-averaged $\rho (s,t)$  and such $u$ inserted in Eq. 
(\ref{rhotau}), then the derivative of the imaginary part naturally produces
the derivative of $\rho (t)$. The resulting differential equation for 
$\rho (t)$ was solved. The strongly non-linear $t$-dependence with large 
negative values of $\rho $ in the Orear region was obtained. The more rigorous
approach was also attempted.

These indications suit quite well the results of the fit in the Orear region 
at 7 TeV \cite{drne} where the negative and quite large values of 
$\rho \approx -2.1$ had to be chosen in that region. No such tendency is 
provided directly by Eq. (\ref{rhotau}).

\begin{figure}[h]
%plot0_1.2_err.gnu
 \includegraphics[width=\textwidth]{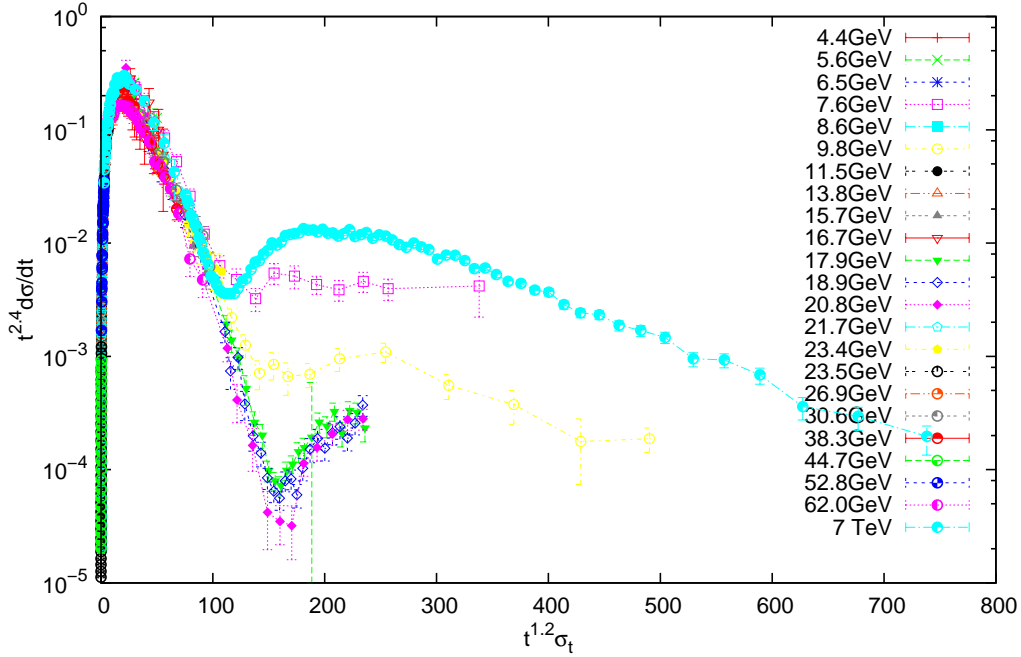}
 \caption{(Color online) The values of $t^{2.4}d\sigma /dt$ for pp-scattering at energies 
$\sqrt s$ from 4.4 GeV to 7 TeV as functions of $t^{1.2}\sigma _t$. 
$t_0=-1$ GeV$^2$ and $\sigma _t$ in GeV$^{-2}$.}
 \label{fit_sigma_tot_1.2}
\end{figure}

The violation of the simple geometrical scaling law (\ref{sl1}) is clearly 
seen in Fig.~\ref{fit_sigma_tot}. In the diffraction cone it is rather well 
satisfied at most energies except the highest one of 7 TeV. In the Orear region 
there is no scaling even at lower energies. We ascribe it to necessary 
modifications of Eq. (\ref{rhotau}). Until now we are unable to propose any 
admissible generalization of Eq. (\ref{rhotau}) at large $t$. 

Nevertheless, we try to modify it at small $t$ in such a way to get better 
scaling inside the diffraction cone even at 7 TeV compared to Fig.~\ref{fit_sigma_tot}.
This can be done by varying the coefficient in front of the linear term in
Eq. (\ref{rhotau}) writing it now as
\begin{equation}
\rho (s,t)=\rho (s,0) \left [1+\frac{1}{a}\frac {\partial \ln {\rm Im}A(s,t) }
{\partial \ln \vert t\vert }\right ].
\label{rhotaua}
\end{equation}
Another way to interpret this modification is to say that all higher order 
terms in $t$-expansion sum to a constant. After the corresponding 
renormalization at $t=0$ it is reduced to Eq. (\ref{rhotaua}).

It can hardly be valid at large $t$ but may help at
low $t$. The equation (\ref{partial}) is easily transformed and the final
prediction is the following scaling law for the differential cross section
\begin{equation}
t^{2a} d\sigma /dt=\omega (t^a\sigma _t).
\label{sla}
\end{equation}

\begin{figure}[h]
%plot0_err.gnu
 \includegraphics[width=\textwidth]{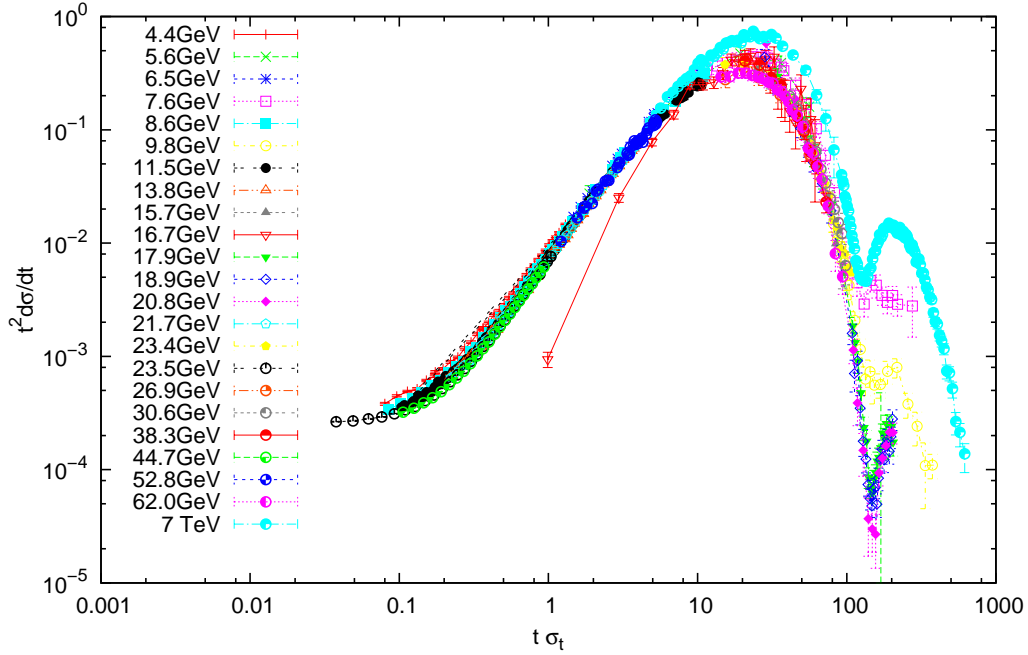}
 \caption{(Color online) The log-log plot of Fig.~\ref{fit_sigma_tot}.}
 \label{fit_sigma_tot_loglog}
\end{figure}

\begin{figure}[h]
%plot0_1.2_err.gnu
 \includegraphics[width=\textwidth]{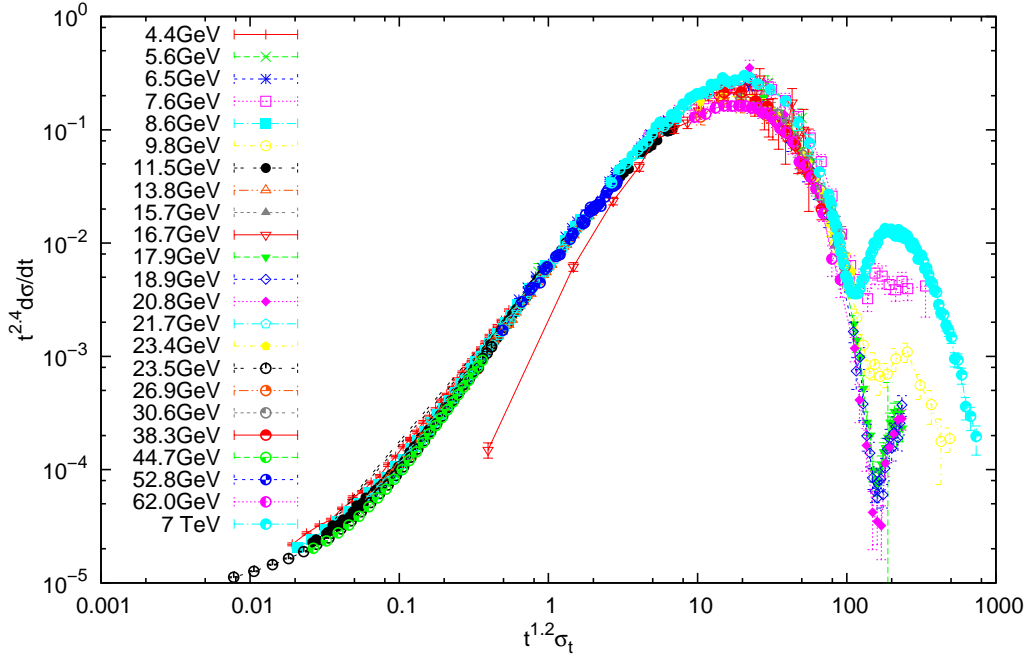}
 \caption{(Color online) The log-log plot of Fig.~\ref{fit_sigma_tot_1.2}.}
 \label{fit_sigma_tot_1.2_loglog}
\end{figure}

We have found that the cross section scales within the diffraction cone at all
available energies in the best way at $a$=1.2. This is shown 
in Fig.~\ref{fit_sigma_tot_1.2}. Thus 
the violation of the geometrical scaling in the diffraction cone has been cured 
with the help of a single parameter $a$. This parameter accounts for the
energy increase of the protons blackness which leads to different energy 
behaviors of $\sigma _t$ and $B$ as mentioned above. The coincidence of maxima 
positions in Fig. 2 implies the relation $\sigma _t\propto B^{1.2}$. That shifts 
the TOTEM maximum to larger values of $t\sigma _t$ in Fig. 1 and allows it to 
fit other maxima at the scale $t^{1.2}\sigma _t$ in Fig. 2. We were able to get 
an approximate scaling in the diffraction cone by some modification of the linear 
in $t$ term only. The somewhat different heigth of the maxima may be ascribed 
to slight variations (oscillations?) of the differential cross section about a 
simple exponent near $t=t_m$ as has been claimed in some experimental studies
\cite{den}. The "effective" parameter $a$ accounts approximately for the opacity
increase with energy. Some other relation (e.g., of the logarithmic shape) is
possible but asks for more elaborated modifications of the Eq. (\ref{rhotau}). 

The region outside the diffraction peak is still not described by 
this law. The scaling violation is noticeable, in particular, even in the shift 
of the positions of the minima closest to the diffraction cone (Its 
parametrization can be found in \cite{bddd}.) not to say about the tails.
That asks for other terms to be added in Eq. (\ref{rhotau}).
Non-linear in $t$ terms would give rise to non-linear modifications of Eq.
(\ref{partial}).

The Figures \ref{fit_sigma_tot} and \ref{fit_sigma_tot_1.2} are redrawn as 
the log-log plots in 
Figs ~\ref{fit_sigma_tot_loglog} and ~\ref{fit_sigma_tot_1.2_loglog}.
Again, the fit with $a$=1.2 is better than with $a$=1.

We have obtained the similar plots both for $a=1$ and $a=1.2$ if $\sigma _t$ is 
approximated by the often used phenomenological dependences $\ln ^2(s/s_0)$ 
and $s^{\Delta}$  with $\Delta$=0.17. Both approximations are not ideally 
suited for fits of the total cross section in the whole energy range and it 
reveals itself in the slight change of scaling shapes. The decline from
the scaling shape, which we do not demonstrate here, is not strong and is 
determined by the accuracy of these approximations. Therefore the fit of 
Fig.~\ref{fit_sigma_tot_1.2} with experimentally known $\sigma _t$ is more 
reliable than those using the above approximations.

To conclude, we have shown that simple geometrical scaling is not fulfilled 
even in the diffraction cone (Figs \ref{fit_sigma_tot} and 
\ref{fit_sigma_tot_loglog}). The physics origin of its violation is that
protons become more black with increasing energy. Nevertheless, the scaling 
law can be generalized to get the agreement in this region of transferred 
momenta so that the differential cross sections fill in a single curve at all 
available energies (Figs \ref{fit_sigma_tot_1.2} and \ref{fit_sigma_tot_1.2_loglog}).
The problem of different tails in the Orear region is related, in our opinion,
to further extensions of the relation (\ref{rhotau}).

%\newpage

{\bf Acknowledgment}

We thank David d'Enterria for the proposal to display Figs 3 and 4 as well,
S.P. Denisov for the reference \cite{den} and S. White for helpful comments.

This work was supported by the RFBR grant 12-02-91504-CERN-a and
by the RAS-CERN program.                                   

%\newpage

\end{document}